\documentclass[runningheads]{llncs}
\usepackage{graphicx}
\usepackage{booktabs}
\usepackage{amsmath,amssymb}
\usepackage{blindtext, xcolor}
\usepackage{comment}
\usepackage{url}
\usepackage[colorinlistoftodos]{todonotes}
\begin{document}
\title{Multi-Domain Adaptation in Brain MRI through Paired Consistency and Adversarial Learning}
\titlerunning{Multi-Domain Adaptation in Brain MRI}
% If the paper title is too long for the running head, you can set
% an abbreviated paper title here
%
\author{Mauricio Orbes-Arteaga**\inst{1,2}\ \and Thomas Varsavsky**\inst{3,1}\ \and Carole H. Sudre \inst{1,4,3}  \and\\ Zach Eaton-Rosen \inst{3,1} \and Lewis J. Haddow \inst{5} \and Lauge S{\o}rensen\inst{2,6,7} \and \\ Mads Nielsen\inst{2,6,7} \and Akshay Pai\inst{2,6,7} \and S\'ebastien Ourselin \inst{1} \and Marc Modat \inst{1}\and \\ Parashkev Nachev\inst{4} \and M. Jorge Cardoso \inst{1}}
\authorrunning{Orbes \& Varsavsky et al.}
% First names are abbreviated in the running head.
% If there are more than two authors, 'et al.' is used.
%
\institute{Biomedical Engineering and Imaging Sciences, King's College London, UK \and Biomediq A/S, Denmark \and  Department of Medical Physics and Biomedical Engineering, UCL, UK \and Institute of Neurology, University College London, UK \and Chelsea and Westminster Hospital NHS Foundation Trust, UK \and Cereriu A/S, Denmark \and Department of Computer Science, University of Coppenhagen, Denmark
\let\thefootnote\relax\footnotetext{** equal contributions}
}
\maketitle              % typeset the header of the contribution
\begin{abstract}
Supervised learning algorithms trained on medical images will often fail to generalize across changes in acquisition parameters. Recent work in domain adaptation addresses this challenge and successfully leverages labeled data in a source domain to perform well on an unlabeled target domain. Inspired by recent work in semi-supervised learning we introduce a novel method to adapt from one source domain to $n$ target domains (as long as there is paired data covering all domains). Our multi-domain adaptation method utilises a consistency loss combined with adversarial learning. We provide results on white matter lesion hyperintensity segmentation from brain MRIs using the MICCAI 2017 challenge data as the source domain and two target domains. The proposed method significantly outperforms other domain adaptation baselines.
\end{abstract}
\section{Introduction}
\label{intro}

%  - Introduce the need for domain adaptation in medical imaging. \\
In medical imaging, fully automated tools using deep learning techniques are increasing in popularity for numerous clinical tasks, including image segmentation, image classification and instance counting \cite{Miotto}. %todo references
Among these tools, deep learning frameworks exhibit excellent performance (often described as `superhuman') when applied on images  drawn from the same distribution (scanner type, parameters, patient pool etc.) as the one used in training the model. However, the performance may deteriorate drastically when the algorithm is applied in previously unseen domains. This performance gap is a critical barrier to the safe implementation and widespread adoption of these techniques in clinical practice. 

The process of adapting a model from a `source' domain to a target domain is called `domain adaptation'. Successful methods have included: 
\begin{enumerate}
    \item Training with a small number of labeled examples from the target domain. While this solution is theoretically straightforward, its practical use is limited as it requires additional labelling on the target domain. 
%TODO extend 
% COMMENT Maybe talk about one shot systems.
    \item Embedding the imaging data in a latent space. This latent space is learnt so as to ignore domain-specific features (e.g. contrast), while retaining domain-invariant features (pathology). Adversarial approaches have been proposed to address this angle in the context of lesion segmentation~\cite{kamnitsas2017unsupervised}and heart structure segmentation between MR and CT~\cite{dou2018unsupervised}. In both cases, the adversarial training is used to make the latent space as uninformative as possible about the domain the images come from. 
    % \item Iteratively including unlabeled data to the training. The trained framework is used to label newly incoming data that is the included in the training set~\cite{Bai2017} 
    \item Semi-supervised methods use a model trained on a small number of labeled examples to provide pseudo-labels for unlabeled data, which is then trained on. The model-fitting and updating semi-supervised labels can be seen as a form of expectation maximisation and has been used in medical imaging~\cite{Bai2017}. 
    \item Enforcing output robustness to input perturbation. Recent methods have exploited the property that the distribution of predictions should be invariant to small perturbations on the input data. This observation can be expressed as $p(y |x)\approx p(y | \Tilde{x})$, where $\Tilde{x}$ is an augmented/perturbed version of $x$. The enforcement of this property has the additional advantage of limiting the unwanted behaviour of drastic output change for minimal input perturbations, which can be seen as improving robustness. 
    % what is the following sentence meant to mean? 
    For instance Perone et al. \cite{perone2019unsupervised} proposed a teacher-student framework ensuring consistency between the outputs when passing to the student an augmented version of the unlabeled input of the teacher, that is similarly augmented afterwards. 
    %\item Hybrid iterative training set enlargement and output robustness enforcement. % TO CONTINUE...
    
    %todo: think if this should include intuition wrt adversarial training, i.e. do we mention that small perturbations can be responsible for large output differences? 
    % COMMENT Then add something in the first item like: "but these solutions can lead to chaotic results in which a very small input perturbation can lead to dramatic output changes"
\end{enumerate}{}
\linepenalty=1000
% COMMENT: You named method 3 before as UDA. Why 2 and 3 now?
Methods 2 to 4 fall under the purview of `Unsupervised Domain Adaptation' (UDA), as does the presented work. In general, UDA does not rely on labeled training examples from the desired target domain. This is especially desirable in medical imaging, where labelling is time-consuming and highly variable, and the `domain' depends on scanner manufacturer, acquisition protocol and reconstruction strategy.
The augmentations required to create the perturbed input data can either be generic (geometric or contrast operations) or application-specific. In the context of medical imaging, the latter includes physics-based image augmentation, synthetic bias field addition or registration-based approaches~\cite{zhao}. These methods lean on domain-specific knowledge to generate plausible transformations.

% We propose a UDA pipeline applied to the segmentation of white matter hyperintensities (WMH) without domain synthetic generation. Instead we enforce the output consistency between the results obtained on two separate acquisitions per subject: a 2D and a 3D FLAIR sequence. We aim to 1) segment WMH lesions on completely unlabeled examples and 2) to make these predictions similar between the 2d and 3d cases. In other words, we regularize the fitting by explicitly promoting similarity in the labels generated by 2d or 3d data. After an overview of the proposed approach in section \ref{methods}, we present the experiments which show that our proposed method leverages the unlabeled data to produce more consistent lesion segmentations across all domains.

We propose a UDA pipeline applied to the segmentation of white matter hyperintensities (WMH) which introduces a paired consistency (PC) loss which guides the adaptation.  The proposed (PC) method enforces the output consistency between the results obtained on two separate acquisitions per subject: an in-plane and a volumetric FLAIR sequence. We aim to 1) segment WMH lesions on completely unlabeled examples and 2) to make these predictions similar between the in-plane and volumetric cases. In other words, we regularize the fitting by explicitly promoting similarity in the labels generated by each FLAIR acquisition. This adaptation method was supplemented with an adversarial loss in order to prevent the model from getting stuck in bad local minima. After an overview of the proposed approach and its variants in section \ref{methods}, we present the experiments which show that our proposed method leverages the unlabeled data to produce more consistent lesion segmentations across all domains.

\section{Methods}
\label{methods}
The proposed training strategy for domain adaptation occurs in two phases. In the first phase the network is trained only on labeled data until convergence. During the second phase of the training, the paired unlabeled data is presented in addition to the labeled data and a consistency term is added to the loss function. This consistency term is inspired from the loss proposed by Xie et al \cite{xie2019unsupervised} that aims at minimizing the Kullback-Leibler divergence $\mathcal{D}_{KL}$ between the output probability distribution  $y$ when conditioned on the unlabeled input $x$ from the set $U$ or its augmented countrapart $\hat{x}$ drawn from $q(\hat{x}|x)$. 
\begin{equation}
    \min_{\theta} \mathcal{L}_{PC} = \underset{x \in U}{\mathbb{E}} \underset{\hat{x} \sim q(\hat{x}|x)}{\mathbb{E}} [ \mathcal{D}_{KL} (p_{\Tilde{\theta}}(y |x) ||  p_{\theta}(y |\hat{x}))]
\end{equation}
% The parameters of the network are $\theta$ and $\Tilde{\theta}$ is a copy of these parameters which we do not differentiate through.\\

We adapt this method to the segmentation task by using the dice loss \cite{milletari2016v} instead of the KL divergence. In the following, we denote as $y_l$ the labeled ground truth, $\hat{y}_l$ the prediction over labeled images, $\hat{y}_u$ the prediction over unlabeled input and $\hat{y}_{\hat{u}}$ the prediction over its augmented/paired counterpart. The losses used in our framework are thus expressed as follows:\looseness=-1

%In order to adapt this method from semi-supervised learning to domain adaptation the unlabeled set $U$ becomes the target set $T$. \\

\begin{equation}
    \mathcal{L}_S = dice(\hat{y_l}, y), \quad
    \mathcal{L}_{PC} = dice(\hat{y_u}, \hat{y}_{\hat{u}}), \quad 
    \mathcal{L}_{tot} = \mathcal{L}_S + \alpha \mathcal{L}_{PC}
\label{eq:2}
\end{equation}

We trained networks using $\mathcal{L}_{tot}$ as specified in (\ref{eq:2}) and denote them as \textit{PC}. These networks $f_{\theta}(h|x)$ produce a feature representation $h$ from which $\hat{y}$ is calculated.
%%%---%%%%%%---%%%%%%---%%%%%%---%%%%%%---%%%%%%---%%%%%%---%%%%%%---%%%%%%---%%%%%%---%%%%%%---%%%%%%---%%% 

\begin{figure}
    \centering
    \includegraphics[scale=0.12]{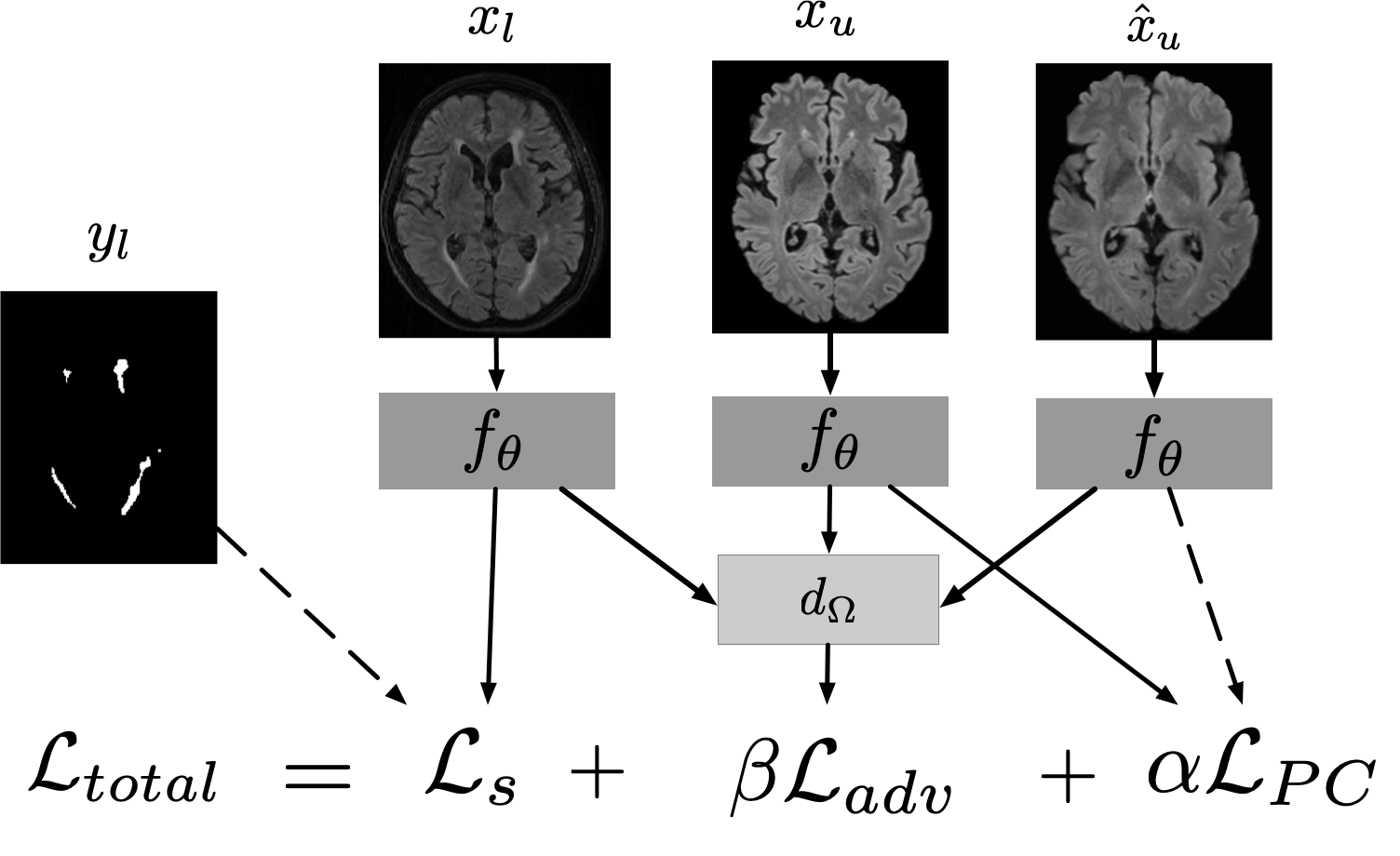}
    \caption{
    Diagram of proposed method. At training time, $x_u$, $x_l$ and $y_l$ are supplied to the network. $x_u$ is an image from the unlabeled target domain  and $\hat{x}_u$ is the result of applying some augmentation function to $x_u$.
    A labeled image, $x_l$, is passed through the network, $f_\theta$ before combining with a label $y_l$ to form the segmentation loss, $\mathcal{L}_s$. The image representations are fed to a domain discriminator $d_\Omega$ which attempts to maximise the cross-entropy between predicted domain and actual domain,  $\mathcal{L}_{adv}$.  Finally, similarity is promoted between the network predictions on $x_u$ and $\hat{x}_u$ using $\mathcal{L}_{PC}$.
    }
    \label{fig:my_label}
\end{figure}

\textbf{Preventing trivial solutions:}  
Early in our experiments, we encountered a specific degenerate solution: our network was able to produce one solution for source images (a good lesion mask) while producing a trivial result on the target domain (in this case, a mask of the foreground). This meant that there was good agreement between in-plane and volumetric FLAIRs because they simply segmented foreground --- ignoring the lesions altogether. This means that the network was identifying the domain of the images and using this to inform its solution: undesirable behaviour. 
%If $h$ contains information about the domain and the network has sufficient parameters then it will be able to predict lesions when an image from the source domain is presented whilst predicting something else on the target domain which maximizes dice. These degenerate solutions were observed in practice. The whole brains in the target domain were being labeled as lesion by the network causing a high dice agreement between them (low loss). 
%%%%%%---%%%%%%---%%%%%%---%%%%%%---%%%%%%---%%%%%%---%%%%%%---%%%%%%---%%%%%%---%%%%%%---%%%%%%---%%%%%%---%%%
We introduced an additional adversarial term to avoid these `solutions'. Inspired by the domain adversarial literature (methods 2 in section~\ref{intro}) we propose an adversarial loss to minimize the amount of information about domain contained in $h$. We introduce a discriminator $d_{\Omega}$ which takes $h$ as input and outputs a domain prediction $\hat{d}$. The adversarial loss, $\mathcal{L}_{adv}$ is given by the cross-entropy, 
$\mathcal{L}_{adv} = - \sum_{i=1}^n \mathcal{L}^{i}_{ce}(d_i, \hat{d_i})$
where $n$ is the number of domains, $\mathcal{L}^{i}_{ce}$ is the multi-class cross entropy loss, $d$ is a one-hot encoded vector of the domain label and $\hat{d}$ is the model's domain prediction as in \cite{schoenauer2019multi}.  We use a gradient reversal layer as in \cite{kamnitsas2017unsupervised} in order to minimize $L_{tot}$ whilst maximizing $L_{adv}$. Figure \ref{fig:my_label} presents the diagram of the proposed method with the combination of different losses, where $\beta$ controls the strength with which the model is adapting its features whereas $\alpha$ controls the weights the consistency effect.
\begin{comment}
\textcolor{red}{COMMENT CAROLE: xhatu and xu are exactly the same image is this normal?}
\end{comment}

\textbf{Augmentation:} In~\cite{xie2019unsupervised} the authors suggest various properties of augmented samples necessary for performing Unsupervised Data Augmentation. Samples should be realistic, valid (meaning they should not alter the underlying label), smooth, diverse and make use of targeted inductive biases (domain knowledge). In the absence of sufficiently realistic augmentation functions we use paired scans which are considered as augmented samples. However, taking them as they are makes for a discrete augmentation function with discontinuous jumps. In order to encourage continuity we used a large range of augmentations on the paired data, including generic geometric transformations and MR specific non-geometric transformations. Geometric augmentations were sampled independently and combined as one affine transform, using random rotations (all axis ranging from -10 to 10 degrees), random shears ($[0.5,0.5]$) and random scaling ($[0.75,1.5]$). For the non-geometric augmentations we applied k-space motion artefact augmentation as described in~\cite{shaw2019mri} and bias field augmentation as implemented in~\cite{gibson2017niftynet}. We measure how useful these additional augmentations are in our experiments.

\section{Experiments and Results}

\subsubsection{Data:}
In this work we focus on white matter hyperintensity segmentation. The data comes from two separate studies. As a source domain we use  the White Matter Hyperintensity challenge data presented in MICCAI 2017 \cite{kuijf2019standardized}. 
%This dataset provides ground truth annotation for 60 subjects from memory clinics with acquisitions equally divided across three scanners: 3T Philips Achieva, 3T Siemens TrioTim and 3T GE Signa HDxt. The Philips and Siemens scanners used an in-plane FLAIR acquisition with $0.96 \times 0.95 \times 3.00$ $mm^3$ voxel sizes whilst the GE scanner used a 3D FLAIR sequence of resolution $0.98 \times 0.98 \times 1.2$ $mm^3$. 
 The other dataset was used as target domain and comes from a sub-study within the Pharmacokinetic and Clinical Observations in PeoPle over fiftY (POPPY)~\cite{haddow2019magnetic}. In this study two different FLAIR sequences were acquired during the same MR session for all 72 subjects on a Philips 3T scanner. The in-plane FLAIR was an axial acquisition with 3mm slice thickness and 1mm\textsuperscript{2} planar resolution (Repetition time (TR) 8000ms, Inversion time (TI) 2400 ms and echo time (TE) 125 ms) while the volumetric FLAIR was of resolution $1.04 \times 1.04 \times 0.56$ mm\textsuperscript{3} (TR=8000ms TI=1650ms TE=282ms). Both images were rigidly coregistered to the 1mm\textsuperscript{3} T1 sequence acquired during the session. All individuals were male with mean age of $59.1 \pm 6.9$ yrs, including HIV-positive subjects and population-matched controls.
\subsubsection{Implementation details and training:}
The MICCAI Challenge dataset was split with a train:validation:test assignment of 40:10:10 subjects. For the POPPY dataset, the split was 38:15:20.

Training was done using 2d axial slices of size 256$\times$256 with inference carried out by concatenating the predictions across all slices to form a 3d volume. The segmentation network uses the U-Net architecture~\cite{ronneberger2015u} with depth of 4 and a maximum number of filters of 256 at the deepest layer, with ReLU as the activation function. Initial training on the MICCAI dataset only was performed using the Adam optimizer with an initial learning rate $10^{-3}$ and a learning rate decay schedule decaying with $\gamma=0.1$ ($\gamma$ is a multiplicative factor of learning rate decay) at epoch 300 and 350. The validation set is used for early stopping, thus the  baseline model takes the network configuration at the epoch where it showed the highest accuracy on the validation set.
All adaptation models and adversarial models were initialized with the weights of this trained baseline model. 

The choice of $\alpha$ parameter balancing the segmentation and the consistency loss in the domain adaptation runs proved to be important. Generally, high values of $\alpha$ led to degenerate solutions, where predictions on the target dataset were no longer capturing lesions. Since scheduling a slowly increasing $\alpha$ did not help, $\alpha$ was fixed at 0.2 in all experiments.\looseness=-1

In case of an adversarial setting, empirical assessment of the best choice of architecture for the discriminator led to the following choice: four 2D convolutional layers with a kernel size of 3$\times$3 and a stride of 2 followed by batch normalisation and leaky ReLU activation. The number of output channels is 4 to begin with and doubles at each layer to a total of 32. Finally, there are three fully connected layers with output sizes of 64, 32, and 2 with relu activations and dropout applied ($p=0.5$).

%\todo{COMMENT CAROLE: It seems a bit strange to name the proposed method UDA since all methods 2 to 4 from the intro fall into the UDA definition. Can we change the names into: Unet, Unet+Aug, Unet+Aug+Adversarial, Unet+Mean Teacher, Unet+Adversarial, Unet+Cons+Adversarial+Aug, Unet+Cons...}

\subsubsection{Points of comparison:}
In order to assess the relevance of the proposed paired consistency, we compared the proposed PC with adversarial setting and augmentation (PC+Adv+Aug) to the version without adversarial setting (PC+Aug) and the simplest version removing also the augmentation (PC). In addition, we trained classical UDA methods with a mean-teacher framework (MT) as well as the adversarial setting without PC with (Adv+Aug) and without augmentation (Adv). Finally we compared to the baseline U-Net model trained only on the MICCAI dataset with (Baseline+Aug) and without (Baseline) augmentation.
% Different flavours (with or without augmentation and adversarial component) of the proposed paired consistency (PC) training were compared to other classical UDA methods (Adversarial and mean teacher with or without augmentation (Aug) ) and to the baseline training.

For the final results table checkpoints were chosen for each of the experiments by looking at the performance across the validation set.

%\subsection{Comparison}
% TO DO CSF COMMENT: Specify against what you trained/ competed

\subsubsection{Reported metrics:}
\label{sec:reported_metrics}

As a first metric of consistency, we compute the Dice score overlap between the two volumes. However, high dice agreement may arise without predicting lesions, for instance with the segmentation of foreground or of another anatomical structure.  Such degenerate solutions can indeed occur as the consistency term in the loss can be minimized for any consistent prediction between volumes. 
 As there are no lesion segmentations for the POPPY dataset, we use the known association between age and white matter hyperintensity load reported for this dataset~\cite{haddow2019magnetic} as surrogate evaluation that the segmented elements are lesions. The effect size is a useful metric for determining whether the lesion loads predicted by the various models agree with the reported literature. For the eight compared models, the effect size ranged from 1.2-fold to 1.5-fold increase in lesion load normalized by total intracranial volume per decade. This compares well with the reported effect size on the POPPY dataset of 1.4-fold with a 95\textsuperscript{th} confidence interval of $\left[1.0 ; 2.0\right]$. 
% We assume from qualitative results and age effect prediction that the lesion load estimates from the models are realistic, thus we compare our models on their ability to minimize the difference in predictions between the two target domains. In order to compare the models we use the same evaluation metric as was used in \cite{kuijf2019standardized}. 
Predictions from in-plane POPPY and volumetric POPPY were compared using the  dice overlap, the  95\textsuperscript{th} percentile Hausdorff distance measured in mm (HD95), the recall (or sensitivity), the ratio of difference in volume between the two predictions (VD) as was used in \cite{kuijf2019standardized}.

\begin{table}[tp]
  \centering
  \caption{Performance of different methods on the target  (POPPY) and the source domain (MICCAI 2017 WMH Challenge). We report the dice between our models' predictions and the ground truth annotations in the source domain as well as the HD95. The evaluation on target domains is done with the Dice, the HD95, the volume difference (VD) and the recall. A significative rank measure is calculated across all metrics. Results are reported with the format median (IQR) in percentages for all metrics except the HD95 in mm. Best results are in bold andunderlined when significantly better than all others (p$<$0.05 paired Wilcoxon tests).}\label{fig:quantitative}
  \resizebox{\textwidth}{!}{
  
    \begin{tabular}{cccccccc}
    \toprule
          & \multicolumn{4}{c}{POPPY}     & \multicolumn{2}{c}{MICCAI} &  \\
          & Dice  & HD    & VD    & Recall & Dice  & HD    & \multicolumn{1}{l}{Rank } \\
          \midrule
    PC+Adv+Aug    & \textbf{\underline{54.5} (10.6)} & 32.7 (9.8) & \textbf{\underline{15.2} (22.8)} & \textbf{\underline{52.4} (14.4)} & 81.4 (9.6) & 28.5 (8.6) & 2.5 \\
    PC+Aug   & 53.2 (15.1) & 39.2 (15.5) & 25.4 (15.6) & 43.5 (12.5) & 81.6 (15.5) & 18.6 (4.8) & 3.3 \\
    PC    & 50.7 (17.0) & 35.1 (11.9) & 16.6 (21.4) & 43.6 (11.0) & 81.4 (22.6) & \textbf{17.2 (3.6)} & 3.4 \\
    MT    & 48.6 (12.3) & 33.6 (14.8) & 33.7 (19.0) & 40.9 (5.0) & 80.0 (18.2) & 20.0 (7.3) & 4.3 \\
    Baseline+Aug    & 42.8 (14.6) & 34.9 (11.1) & 39.3 (22.3) & 33.5 (12.6) & 80.6 (14.8) & 17.8 (4.9) & 4.9 \\
    Baseline    & 43.0 (16.2) & 33.3 (15.1) & 40.3 (24.8) & 33.3 (14.8) & 81.1 (16.9) & 17.5 (3.3) & 5.6 \\
    Adv    & 41.8 (15.4) & \textbf{32.6 (6.1)} & 25.2 (24.0) & 33.5 (12.7) & \textbf{82.5 (12.0)} & 17.6 (5.2) & 5.7 \\
    Adv+Aug    & 41.4 (16.4) & 36.6 (9.0) & 38.0 (16.0) & 33.6 (13.9) & 81.9 (11.1) & 19.7 (11.0) & 6.3 \\
    \bottomrule
    \end{tabular}%
    }
\end{table}%
The results, gathered in Table \ref{fig:quantitative}, reporting median and interquartile range are ordered according to the average significance ranking, follows the guidelines of the MICCAI Decathlon challenge 2018 \footnote{ \url{http://medicaldecathlon.com/files/MSD-Ranking-scheme.pdf}}. %where HD95 and VD are ranked from low to high and Recall and Dice ranked from high to low. 
\begin{figure}[tp]
    \centering
    \includegraphics[width=0.95\textwidth]{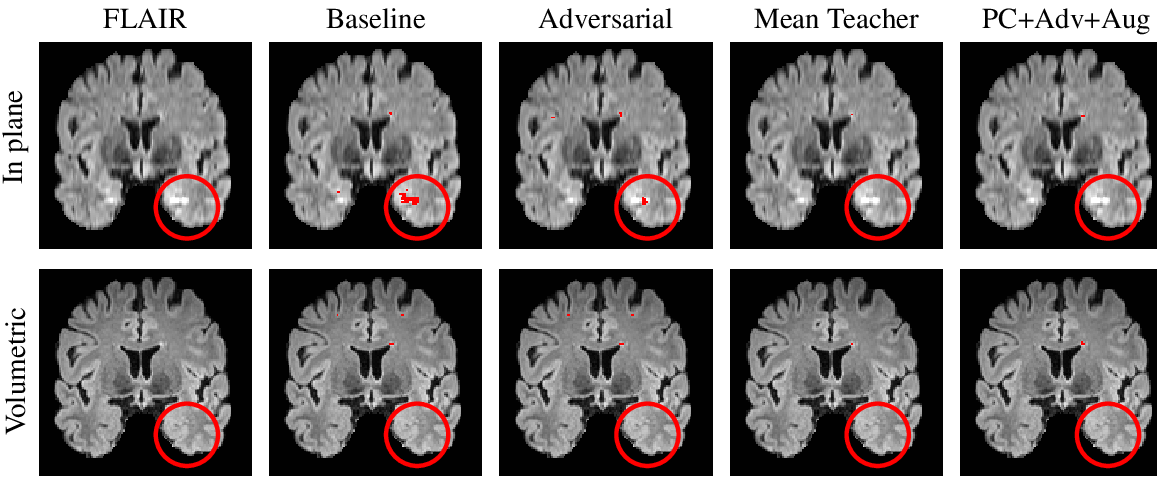}
    \caption{Qualitative results on a single slice from a single subject in the POPPY dataset. The top row shows a slice from the in-plane FLAIR acquisition whilst the bottom row shows a slice from the volumetric FLAIR acquisition. Each column shows a model's predictions on that row's image. This slice is used to highlight an example of an artefact (shown in the red circle) introduced by the in-plane acquisition. The baseline method introduces a false positive in this region whilst the domain adaptation methods perform better at ignoring it. Our approach shows the best in-plane to volumetric agreement.}
    \label{fig:qualitative}
\end{figure}

\section{Discussion}
In this work, we presented a novel method of performing unsupervised multi-domain adaptation. A pretrained model from one domain is retrained on paired unlabeled data from two target domains, encouraging consistent predictions. The proposed approach was evaluated against existing UDA strategies including representation learning approaches using domain adversarial training~\cite{kamnitsas2017unsupervised}, and the `Mean Teacher' algorithm for unsupervised domain adaptation~\cite{xie2019unsupervised} as well as an unsupervised baseline for WMH segmentation. Overall, our method was able to produce more consistent predictions across two target domains while retaining similar performance on its original training domain. 
More specifically, adaptation techniques optimizing pairwise consistency not only outperformed baseline models not benefitting from any adaptation but also adversarial strategies.
Furthermore, it appeared that the PC method while closest to the mean teacher algorithm, outperformed this approach potentially thanks to differences in the optimisation strategies. Understanding the reasons for these differences also reported by \cite{xie2019unsupervised} could be an interesting avenue of future investigation.  % the approaches are similar in spirit, the differences in optimisation may lead to significant performance differences. 
Regarding the adversarial results, the observed inferior performance suggests that depending on the adaptation problem, the learning of a latent space invariant to domain (as enforced in the adversarial approach) may cause an information loss detrimental to the segmentation task. Additionally, the effects of data augmentation (which normally impacts performance positively) did not provide any benefit in the pure adversarial setting. Specific investigation of the effect of each type of augmentation would be needed to better understand this behaviour. While a pure adversarial setting proved ineffective, best performance across all models was obtained when combining it with our proposed PC strategy as it promoted a good label distribution in our target images. Future work will focus on removing the need for paired data by finding sufficiently realistic augmentation functions. % This highlights not only the utility of adaptation strategies for encouraging generalization of segmentation models across domain but also the superiority of consistency loss over the adversarial loss for domain adaptation. On the other hand, the domain adversarial methods were the lowest ranked. This suggests that projecting to a space that is invariant across domains does not necessarily improve agreement between two segmentations. When training an adversarial network, there is a trade-off to be made between having a latent space that works for segmentation, while not yielding information as to the original domain.  The  2d/3d nature of the data may mean that we lose more information than we normally would as we make the latent space invariant to domain.

%UDA vs mean teacher 
%Our proposed method outperforms the `Mean Teacher' method, which agrees with results found in~\cite{xie2019unsupervised}. This result, while empirically sound, is nonetheless confusing: there does not seem to be a clear reason why `Mean Teacher' performs worse in our work or~\cite{xie2019unsupervised}.

% adversarial training helps with our method 

%we did use a discriminator in our best-performing model to encourage a good label distribution in our target images. Our proposed model with this discriminator term was the best-performing model. In contrast, the absence of the consistency part of the framework led generally to higher training instability in the adversarial setting. 

In conclusion, PC is a promising method to adapt automated image segmentation tools to different scanner manufacturers, MR sequences and other confounds. This adaptation is critical to the clinical translation of these tools notably in the context of scanner upgrades and multicentre trials.

\subsubsection{Acknowledgements}
We gratefully acknowledge the support of NVIDIA Corporation with the donation of one Titan Xp. This project has received funding from the EU H2020 under the Marie Sk\l odowska-Curie grant agreement No 721820, Wellcome Flagship Programme (WT213038/Z/18/Z) and Wellcome EPSRC CME (WT203148/Z/16/Z). Carole H. Sudre is supported by AS-JF-17-011 Alzheimer's Society Junior Fellowship.

\bibliographystyle{splncs04}
\bibliography{main}
\end{document}